# Large-area, freestanding single-crystal gold of single-nanometer thickness

*Chenxinyu Pan[1], Yuanbiao Tong[1], Haoliang Qian[2], Alexey V. Krasavin[3], Jialin Li[1], Jiajie Zhu[1], Yiyun Zhang[2], Bowen Cui[1], Zhiyong Li[1,4], Chenming Wu[1], Zhenxin Wang[1], Lufang Liu[1], Linjun Li[1,4], Xin Guo[1,4], Anatoly V. Zayats[3,\*], Limin Tong[1,5,\*] and Pan Wang[1,4,\*]*

[1]Interdisciplinary Center for Quantum Information, State Key Laboratory of Extreme Photonics and Instrumentation, College of Optical Science and Engineering, Zhejiang University, Hangzhou 310027, China

[2]Interdisciplinary Center for Quantum Information, State Key Laboratory of Extreme Photonics and Instrumentation, ZJU-Hangzhou Global Scientific and Technological Innovation Center, Zhejiang University, Hangzhou 310027, China

[3]Department of Physics and London Centre for Nanotechnology, King's College London, Strand, London WC2R 2LS, UK

[4]Jiaxing Key Laboratory of Photonic Sensing & Intelligent Imaging, Intelligent Optics & Photonics Research Center, Jiaxing Research Institute Zhejiang University, Jiaxing 314000, China

[5]Collaborative Innovation Center of Extreme Optics, Shanxi University, Taiyuan 030006, China

\*Corresponding authors

E-mail: a.zayats@kcl.ac.uk

E-mail: phytong@zju.edu.cn

E-mail: nanopan@zju.edu.cn

**Abstract**

**Two-dimensional single-crystal metals are highly sought after for next-generation technologies. Here, we report large-area (>$10^4$ µm$^2$), single-crystal two-dimensional gold with thicknesses down to a single-nanometer level, employing an atomic-level-precision chemical etching approach. The ultrathin thickness and single-crystal quality endow two-dimensional gold with unique properties including significantly quantum-confinement-augmented optical nonlinearity, low sheet resistance, high transparency and excellent mechanical flexibility. By patterning the two-dimensional gold into nanoribbon arrays, extremely-confined near-infrared plasmonic resonances are further demonstrated with quality factors up to 5. The freestanding nature of two-dimensional gold allows its straightforward manipulation and transfer-printing for integration with other structures. The developed two-dimensional gold provides an emerging platform for fundamental studies in various disciplines and opens up new opportunities for applications in high-performance ultrathin optoelectronic, photonic and quantum devices.**

Low-dimensional metals have attracted extraordinary research interest because of their unconventional physical, chemical and mechanical properties arising from the reduced dimensionality[1-4]. Among them, ultrathin (few-nanometer thick) two-dimensional (2D) gold, which combines advantages of quantum effects in its electric and optical properties[5-7], plasmon-enabled extreme light confinement[8-11], high optical transparency and excellent chemical stability, is highly desired for the realization of metal-based ultrathin optoelectronic, photonic and quantum devices[6,8-10,12-14]. To date, various wet-chemical approaches, such as seed-mediated synthesis[15-19], polyol reduction method[20-23], 2D template-directed synthesis[24-31], methyl orange-assisted synthesis[32,33] and others[34-36], have been developed for the fabrication of freestanding 2D gold with high crystalline quality and thicknesses down to a sub-nanometer scale[28,32]. However, due to the proportional increase in the thickness and lateral size with the growth time, it is extremely challenging to fabricate ultrathin 2D gold with a large area. For example, for 2D gold with a sub-5-nm thickness, the lateral size is usually at a sub-micrometer scale[19,24,28,29,31,32,35], which imposes restrictions for most applications. Moreover, it is difficult to precisely control the thickness of 2D gold with wet-chemical approaches, while it is highly desired due to the extreme sensitivity of its electric and optical properties to the thickness because of the quantum confinement or electron surface scattering effects[5,6,10,13]. Ultrathin gold films on dielectric substrates have recently been obtained with deposition approaches using adhesive/seeding layers, such as metals (e.g., copper)[10,13,37] and organosilane monolayers[38], to reduce the percolation threshold of gold. However, as-fabricated gold films have a granular polycrystalline structure that can affect their performance in many applications (e.g., due to the electron scattering losses introduced by surface roughness and grain boundaries[39-42]). In addition, it is difficult to detach them from the substrates due to the existence of adhesion/seeding layers, which greatly limits their flexibility for fundamental studies and applications (e.g., integration with other structures and devices).

Here, we introduce an atomic-level-precision etching (ALPE) approach to circumvent the lateral size-thickness relation in wet-chemical approaches, enabling the fabrication of large-area, freestanding single-crystal 2D gold with thicknesses down to a single-nanometer level. Large-area atomically-smooth gold flakes (Supplementary Fig. 1) were first synthesized on a substrate (e.g., mica)[22,42] as the starting structures. Then, they were immersed into a cysteamine solution to initiate chemical etching. During this process, gold atoms on the surface of the gold flakes were etched monolayer-by-monolayer via the formation of soluble gold-thiolate complexes to transform them into substrate-supported 2D gold flakes (2DGFs, Fig. 1a). Figure 1b presents optical micrographs of the etching of a gold flake with its thickness decreasing from 32 to 1.9 nm (accompanied by a visible increase in transmission), while its lateral size remaining almost unchanged. The etching rate was estimated to be ~0.2 nm/min (Fig. 1c) and is dependent on the concentration of the cysteamine. This indicates the crucial ability to control the thickness of the 2DGF at an atomic-level precision, which is difficult for conventional wet-chemical approaches and is of great importance for the precise engineering of properties of the 2DGF[5,6,10,13]. Because of the absence of adhesive layers, as-fabricated 2DGFs are freestanding, as evidenced by the optical micrograph of a 2.8-nm-thick 2DGF folded on itself (Fig. 1d).

The lateral size of the 2DGF is determined by the size of the initial gold flakes, which can be at a 100s-μm scale (Supplementary Fig. 2). Its area is at least five orders of magnitude larger than that of 2D gold of similar thicknesses obtained with other methods[24,29]. Such a large area makes it possible to exploit 2DGFs for applications in fundamental studies and the construction of ultrathin metal-based structures and devices. The ALPE approach can be applied to spatially localized areas to create micro/nanostructures (see Supplementary Fig. 3). As an example, Figure 1e shows an optical micrograph of a gold flake locally etched to implement a concentric ring pattern, whose thickness is about 10 nm less than that of the surrounding area (Fig. 1f). The ALPE approach can also be applied for the fabrication of 2D silver and copper flakes (Supplementary Fig. 4), showing applicability of this approach to other metals.

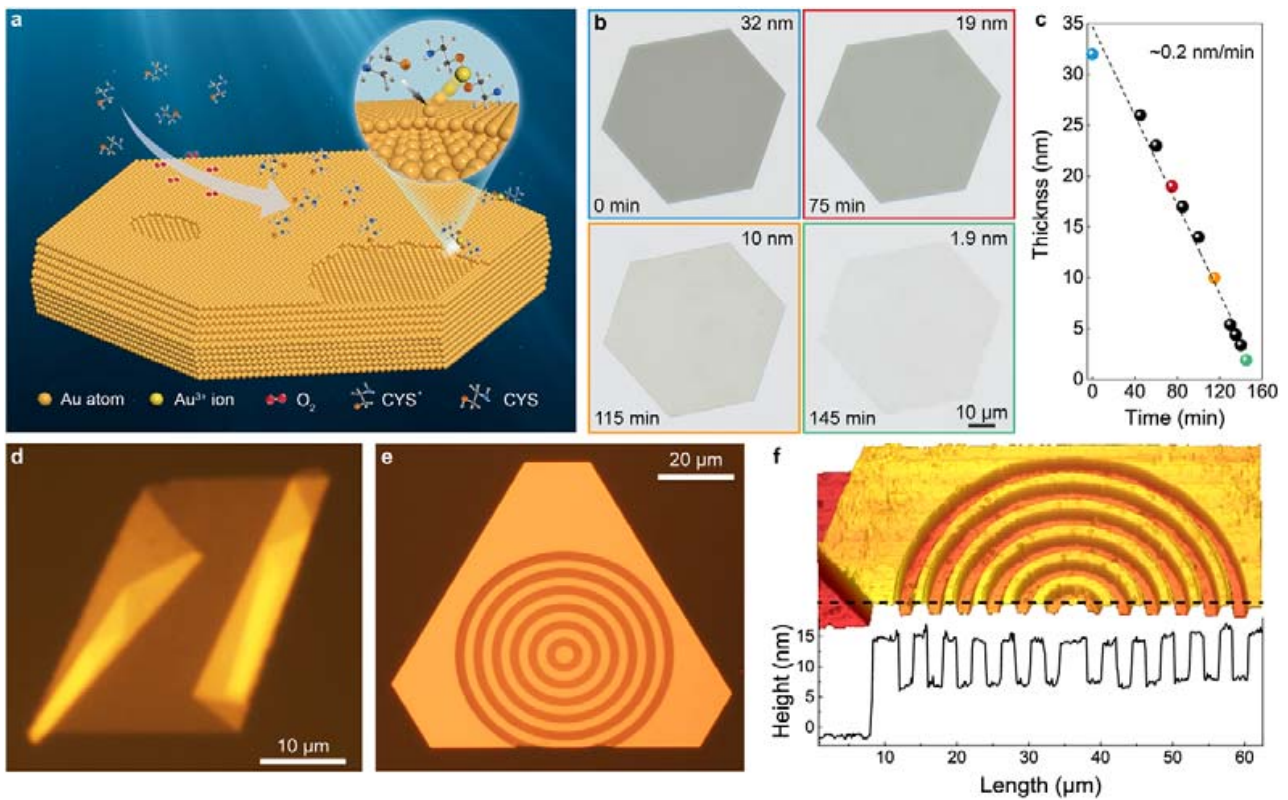

**Figure 1 | Fabrication of 2DGFs. a**, Schematic illustration of the ALPE approach for fabricating 2DGFs. **b**, Optical transmission micrographs of a gold flake taken at various etching times. **c**, Measured thickness of the gold flake as a function of the etching time. **d,e**, Optical reflection micrographs of a 2DGF folded on itself (**d**) and a gold flake locally etched to implement a concentric ring pattern (**e**). **f**, AFM image of the etched flake in **e**. A line scan along the indicated dashed line is also shown.

Atomic force microscopy (AFM) and transmission electron microscopy (TEM) were used to characterize the surface morphology and crystalline structure of as-fabricated 2DGFs. The AFM image of a typical 2DGF (Fig. 2a) shows that it has a thickness of 1.9 nm and an atomically smooth surface with a root-mean-square roughness of ~0.3 nm. This reveals the uniform atomic monolayer-by-monolayer etching characteristic of the approach that can retain the excellent surface quality and thickness uniformity of the initial flake, which is important for obtaining large-area 2DGFs. The high-resolution TEM images of the surface of a 3.7-nm-thick 2DGF (Fig. 2b) show highly periodic lattice fringes with an interplanar spacing of ~0.24 nm (inset). The single-crystal property of the 2DGF was confirmed by the electron diffraction pattern (Fig. 2c), showing a hexagonal close-packed structure with a <0001> crystal orientation. Figure 2d (and Supplementary Fig. 5) further presents a cross-sectional TEM image of a 2DGF (see Methods for details), in which 10 atomic planes of gold can be clearly observed (Fig. 2e). 2DGFs with such a large area is stable under ambient conditions for at least 6 months (Supplementary Fig. 6).

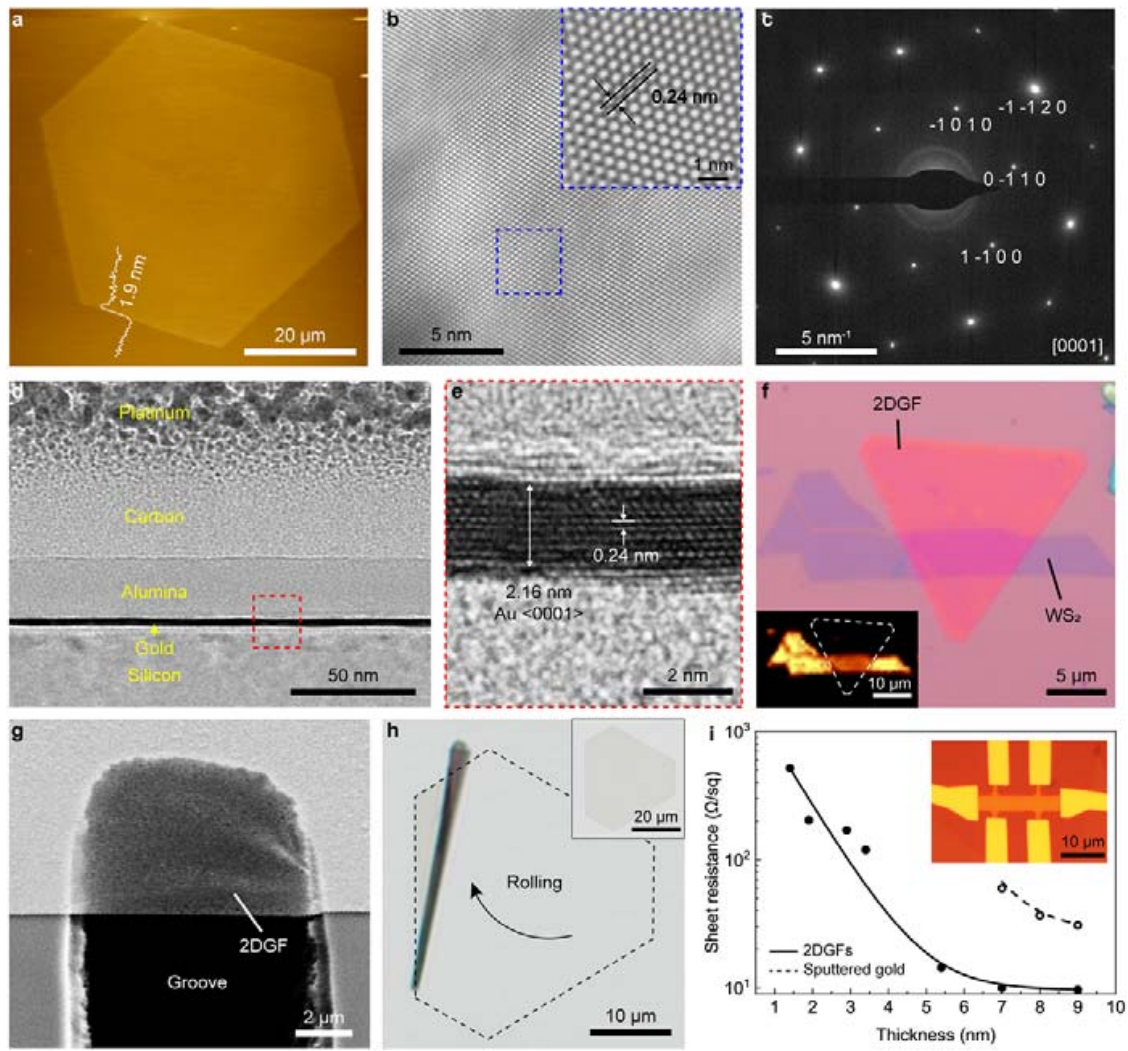

**Figure 2 | Structural and electric properties. a**, AFM image of a 1.9-nm-thick 2DGF. **b**, Planar TEM image of a 3.7-nm-thick 2DGF. Inset, enlarged view of the marked region. **c**, Electron diffraction pattern of the region marked in **b**. **d**, Cross-sectional TEM image of a 2DGF. **e**, Atomic-resolution TEM image of the region marked in **d**. **f**, Optical micrograph of a 2DGF-WS$_2$ heterostructure with the corresponding photoluminescence mapping shown in the inset. **g**, Scanning electron microscopy image of a 3.8-nm-thick 2DGF suspended across a groove. **h**, Optical micrograph of a rolled-up 3.6-nm-thick 2DGF. The dashed line shows the outline of the original 2DGF (inset). **i**, Thickness-dependent sheet resistance of 2DGFs and sputered gold films (the latter is conductive only for thicknesses larger than the percolation thickness of ~7 nm). Inset, optical micrograph of a Hall-bar structure fabricated with a 5.4-nm-thick 2DGF.

Because of their large area, excellent structural quality and freestanding nature, the 2DGFs can be readily manipulated and transfer-printed onto diverse substrates (see Methods and Supplementary Fig. 7), greatly expanding their flexibility for fundamental studies and applications. As an example, three pieces of 2DGFs were sequentially transfer-printed to form a stacked multilayer structure (Supplementary Fig. 8a). In another example, a 2DGF was transfer-printed onto a $WS_2$ monolayer to form an ultrathin metal-semiconductor heterostructure (Fig. 2f). Due to its ultrathin thickness, the 2DGF is transparent, thus, the underlying $WS_2$ monolayer and its photoluminescence excited through the 2DGF (inset) can be clearly seen. Compared to conventional deposition approaches, the transfer printing of 2DGFs provides a gentle and damage-free approach for metal integration, which is especially attractive for delicate materials such as 2D semiconductors and organic molecules[43]. Transfer printing of 2DGFs onto a groove (Fig. 2g) and the curved sidewall of an optical microfiber (Supplementary Fig. 8b) were also demonstrated; either of the systems is difficult to realize with conventional deposition approaches. Furthermore, the 2DGF can be rolled up into a microtube structure with a diameter of ~2 μm (Fig. 2h)[44], showing its excellent mechanical flexibility and the potential for introducing a strain to further engineer its electric and optical properties[45].

Electric properties of the 2DGFs were investigated by using a four-probe approach based on a Hall-bar structure (inset of Fig. 2i, see Methods and Supplementary Fig. 9 for details). With the decrease of the thickness from 9 to 1.4 nm, the sheet resistance of 2DGFs increases from ~9 to 530 Ω per square (Ω/sq) (solid dots, Fig. 2i). By contrast, sputtered gold films are only conductive for thicknesses larger than the percolation threshold of ~7 nm, their sheet resistance increases quickly from ~31 to 61 Ω/sq when the film thickness decreases from 9 to 7 nm (hollow dots, Fig. 2i). The pronounced decrease in the sheet resistance of 2DGFs can be attributed to their high crystal quality and excellent surface smoothness.

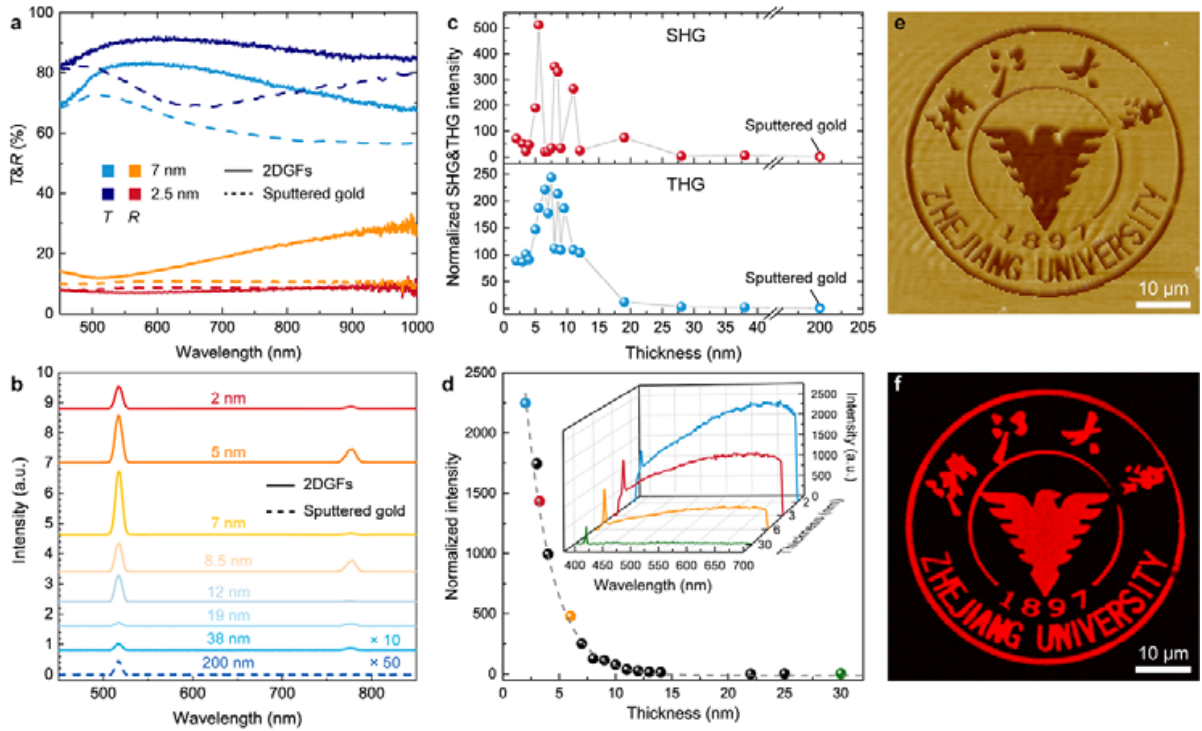

**Figure 3 | Thickness-dependent optical properties. a**, Measured transmittance (*T*) and reflectance (*R*) of 2DGFs (solid lines) and sputtered gold films (dashed lines) with different thicknesses on a mica substrate. **b,c**, Nonlinear emission spectra measured from 2DGFs with various thicknesses and a 200-nm-thick sputtered gold film (**b**) and the thickness dependence of SHG and THG intensities from the 2DGFs normalized with those from the 200-nm-thick sputtered gold film (**c**) under *p*-polarized pulsed laser excitation (1550-nm wavelength, ~140-fs pulse width) at an incident angle of 30°. The spectra in **b** are shifted along the vertical axis for visibility. **d**, Thickness-dependent MPPL intensity from 2DGFs obtained under *p*-polarized pulsed laser excitation (800-nm wavelength, ~100-fs pulse width) at an incident angle of 30°. The inset shows the measured emission spectra from 2DGFs with thicknesses of 2, 3, 6 and 30 nm. **e,f**, AFM image (**e**) of a logo of Zhejiang University locally etched in a 36-nm-thick gold flake (the height contrast is 28 nm) and its MPPL image (**f**) with the signal integrated over a 575–630 nm spectral range.

The decrease in the thickness endows 2DGFs with intriguing optical properties. Transmission and reflection spectra of 2DGFs with different thicknesses were compared to sputtered gold film counterparts (Fig. 3a). Due to the presence of electrically unconnected gold islands in the investigated sputtered gold films, a dip (peak) in transmittance (reflectance) around 650 nm, that red-shifts and widens with increasing thickness, can be observed (dashed lines) as a signature of the excitation of localized surface plasmon (LSP) modes of the gold islands[10]. By contrast, these dips/peaks are absent in the transmission and reflection spectra for 2DGFs with thickness down to 2.5 nm (solid lines), further indicating the excellent continuity of 2DGFs. Benefitting from the outstanding crystalline quality and surface smoothness, 2DGFs have a much higher transmittance compared with that of sputtered gold films with the same thickness. Particularly, for the 2.5-nm-thick 2DGF, the transmittance around 600 nm reaches a value of ~91%.

As the thickness of 2DGFs approaches few nanometers, quantization of the electronic energy in the out-of-plane direction becomes important[5] (Supplementary Fig. 10), which makes a crucial impact on the nonlinear optical response of 2DGFs[6,14,46,47]. Under pulsed laser excitation at 1550-nm wavelength (see Methods and Supplementary Fig. 11), the nonlinear emission spectra from both 2DGFs with various thicknesses and a 200-nm-thick sputtered gold film feature narrow peaks at 775 and 516.7 nm (Fig. 3b), which corresponds to the second-harmonic generation (SHG) and third-harmonic generation (THG) signals, respectively. The SHG and THG intensities from the 2DGFs are much higher than those from the sputtered gold film, and are strongly dependent on the thickness. With the decrease of the thickness, they increase nonmonotonously, exhibiting very sharp oscillations with thickness (Fig. 3c). The SHG and THG intensities are enhanced by ~500 and 250 times for the 2DGFs with thicknesses of 5 and 7 nm, respectively. This can be explained by quantization of the electronic energy levels of the 2DGFs so that the optical transitions excited by photons with a fixed energy between the intersubband levels are in and out of resonance as the thickness changes, resulting in the resonant enhancement[6,46-48]. The resonances occur at different thicknesses of gold for SHG and THG processes and agree well with the simulation results (Supplementary Fig. 12a,b) based on the quantum electrostatic model[6] (see Methods). It is worth noting that the measured thickness-dependent THG is observed on a broad underlying background, which can be attributed to the contribution from interband transitions in the 2DGFs with quantized electronic states[46] (see Methods and Supplementary Fig. 12c). The SHG intensity from 2DGFs under the excitation with 800-nm laser pulses (Fig. 3d, inset) also exhibits the thickness-dependent oscillatory behavior (Supplementary Fig. 13). In addition to the SHG signal, a broad multiphoton photoluminescence (MPPL) background is also observed, identified by typical excitation power dependences (Supplementary Fig. 14). With the decrease of the gold thickness from 30 to 2 nm, the spectrally integrated MPPL intensity increases by ~2200 times (Fig. 3d). The enhancement is so high because in bulk gold MPPL is an extremely inefficient process, as the momentum of the photon is too small to satisfy momentum conservation of the involved intraband transition in the $sp$ conduction band[49]. Benefiting from the thickness-dependent quantization of the energy levels, intersubband transitions in 2DGFs do not require an additional momentum, which greatly boosts the MPPL efficiency[49].

Due to the strong thickness dependence of the nonlinear optical properties, the ALPE approach can be used to locally engineer the optical nonlinearity of gold flakes. As an example, a logo of Zhejiang University was etched into a 36-nm-thick gold flake (Fig. 3e), with its thickness locally modified with a 28-nm step. Under the excitation with 800-nm laser pulses, the logo was observed in the MPPL image taken with a 575–630 nm bandpass filter as a red emission pattern on a completely dark background (Fig. 3f), revealing significantly enhanced MPPL in the thinner logo region compared to a thicker surrounding. This is particularly important for the applications in which high local optical nonlinearity is required without the integration of other, e.g. technologically incompatible, materials.

The excellent crystal quality, atomically-smooth surface and 100s-μm lateral size of 2DGFs enable the realization of low-loss plasmonic nanostructures with extreme optical confinement[8,10,11,50,51]. To implement a typical plasmonic nanostructure supporting LSP resonances, 2DGFs were patterned into nanoribbon arrays using the local etching approach (Fig. 4a). First, nanoribbon arrays with a fixed width ($w$ = 100 nm) and period ($p$ = 3$w$) were fabricated using 2DGFs with thicknesses of 7, 5 and 3 nm. Figure 4b shows an example of an as-fabricated nanoribbon array with a thickness of 3 nm, featuring a smooth surface and having no visible defects. In marked contrast, for nanoribbon arrays fabricated using sputtered gold films (see Methods), the nanoribbons are quite rough and become discontinuous when the thickness is less than ~7 nm (Fig. 4c). As a result, there is no observable LSP mode in their transmission spectra (Fig. 4d, solid lines), however, strong resonance dips due to the excitation of a dipolar LSP mode in each of the nanoribbons can be observed for the nanoribbon arrays fabricated using 2DGFs (Fig. 4d, dashed lines). The resonance dip is red shifted dramatically from 767 to 1115 nm with the decrease of the nanoribbon thickness from 7 to 3 nm by only 4 nm, showing an unprecedented versatility in the plasmonic response as a consequence of the ultrathin thickness. The quality factor for the plasmonic mode of the nanoribbon array with 3-nm thickness is up to ~5, which can be attributed to the greatly reduced electron scattering losses in the ultrasmooth single-crystal 2DGFs[10,50]. The resonance wavelength of the plasmonic modes in the ultrathin nanoribbons can also be tuned by changing their width (Fig. 4e, solid lines). With the decrease of the nanoribbon width from 100 to 75 nm ($t$ = 2.5 nm, $p$ = 3$w$), the resonance dip is blue shifted from 1264 to 951 nm. Numerically simulated transmission through the gold nanoribbon arrays (see Methods) confirms the experimental results (Fig. 4e, dashed lines).

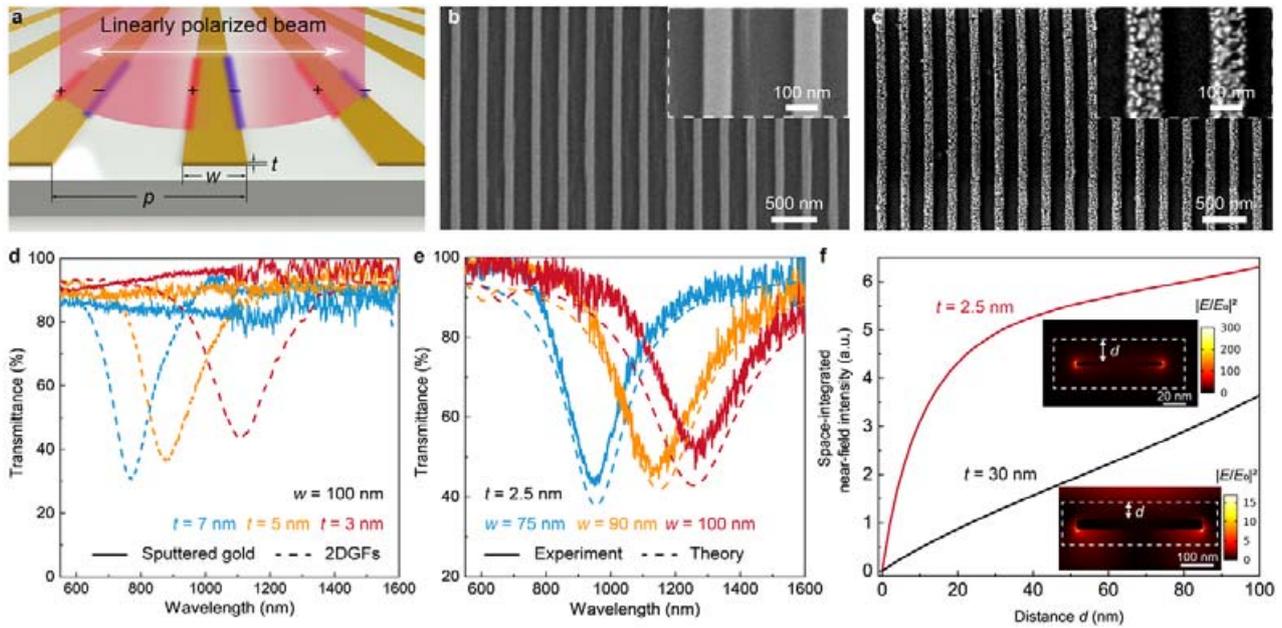

**Figure 4 | 2DGFs for low-loss nanoplasmonics. a**, Schematic illustration of patterning of a 2DGF into a nanoribbon array supporting dipolar LSP modes. **b,c**, Scanning electron microscopy images of nanoribbon arrays fabricated from a 2DGF (**b**) and a sputterred gold film (**c**) with the same thickness of 3 nm. **d**, Transmittance of nanoribbon arrays fabricated with 2DGFs (dashed lines) and sputtered gold films (solid lines) for various nanoribbon thicknesses ($w$ = 100 nm). **e**, Measured (solid lines) and calculated (dashed lines) transmittance of nanoribbon arrays fabricated with 2DGFs for various nanoribbon widths ($t$ = 2.5 nm). **f**, Numerically calculated space-integrated near-field intensities confined within an area extending by a distance $d$ outside the nanoribbons made from a 2.5-nm-thick 2DGF and a 30-nm-thick sputtered gold film. The insets show the corresponding near-field intensity distributions of the nanoribbons having the same resonance wavelength of 970 nm.

Despite similar plasmonic resonances can be obtained in nanoribbon arrays fabricated using sputtered gold films with larger thicknesses (e.g., 30 nm, see Supplementary Fig. 15), nanoribbon arrays with an ultrathin thickness enabled by the 2DGFs, as shown in Fig. 4f, provide attractive advantages including tighter optical field confinement around the nanoribbon (cf., red and black lines), larger local-field enhancement (~20 times higher) and smaller footprint (~5 times reduction in width). Compared with plasmons in graphene and other 2D semiconductors which usually lie at mid-infrared frequencies[52], 2DGFs with a much higher free electron density allows the realization of extremely-confined plasmons operating in the technologically appealing near-infrared spectral range[10,50]. Large-area freestanding 2DGFs, merging low-loss extremely-confined plasmons with the significantly quantum-confinement-augmented optical nonlinearity, allow the access to the regime of extreme light-matter interactions for fundamental studies[8,11,14,51] as well as the realization of ultrathin nanophotonic devices including ultrafast optical modulators[9,10] and high-sensitivity optical sensors[52].

Using an ALPE approach, we have successfully fabricated freestanding single-crystal 2DGFs with dimensions extending over a 100s-μm scale and thicknesses down to a single-nanometer level, which has unique properties including significantly quantum-confinement-augmented optical nonlinearity, plasmon-enabled extreme light confinement, low sheet resistance and high mechanical flexibility. The thickness of 2DGFs can be further pushed down to a sub-nanometer level by optimizing the etching conditions (Supplementary Fig. 16). Large-area freestanding 2DGFs provide an emerging platform for fundamental research in various disciplines, such as physics, electronics, chemistry and mechanics, and promises many unique opportunities in the development of ultrathin optoelectronic, photonic and quantum devices.

## Methods

**TEM characterization.** Planar TEM images were acquired with a Talos F200X G2 high-resolution transmission electron microscope operated at 200 kV. To prepare TEM grids with the samples, a 2DGF was first fabricated on a substrate using the ALPE approach, then a fraction of the 2DGF was picked up from the substrate and transferred onto a copper grid by a fiber taper. For the cross-sectional TEM characterization of 2DGFs, an electron-transparent cross-sectional lamella of the sample was prepared as follows. Firstly, a 2DGF was fabricated on a silicon substrate using the ALPE approach, followed by the deposition of a layer of $Al_2O_3$ (~30 nm in thickness) onto it using atomic layer deposition (SENTECH SI ALD) at 100 °C[42]. Secondly, layers of carbon (~50 nm in thickness) and platinum (~70 nm in thickness) were further sputtered on the sample for protecting it. Then, a cross-sectional lamella of the sample with a thickness less than 50 nm was obtained using a focused-ion-beam system (Helios G4, Thermo Scientific). Finally, the lamella was transferred onto a copper grid, and imaged using a Talos F200X G2 high-resolution transmission electron microscope operated at 200 kV.

**Transfer printing of 2DGFs.** A polydimethylsiloxane (PDMS)-assisted transfer printing approach was used to pick up a 2DGF from a substrate and transfer print it onto targeted substrates or structures. As schematically shown in Supplementary Fig. 7, to facilitate the pick up of the 2DGF from the substrate, water vapor was first evaporated onto a small patch of PDMS film to form numerous micrometer-scale water droplets. Then, under an optical microscope, the PDMS film was pressed onto the targeted 2DGF, and slowly withdrawn backward to pick it up. Finally, the 2DGF attached to the PDMS film was aligned and made a full contact pricisely to the targeted substrate or structure with a three-dimensional moving stage. The 2DGF was left on the substrate or the structure after a careful withdrawal of the PDMS film.

**Measurement of sheet resistance.** A four-probe approach based on a Hall-bar structure was used to investigate the electrical properties of 2DGFs (as schematically shown in Supplementary Fig. 9). After the fabrication of a 2DGF on a 300-nm-thick silica-coated silicon substrate using the ALPE approach, the local etching approach was used to pattern the 2DGF into the Hall-bar structure. Then, EBL was used to fabricate a mask for the deposition of metal contact electrodes (Cr/Au: 5/80 nm), followed by a lift-off process. An optical microscopy image of an as-fabricated Hall-bar structure is shown in the inset of Fig. 2i. The samples for sputtered gold films were fabricated as follows. EBL was first used to define masks of the Hall-bar structures on a 300-nm-thick silica-coated silicon substrate. Secondly, to avoid the use of metallic adhesion layers (such as Cr, Ti) that can affect the electrical property of the gold films, the exposed region of the masks was functionalized with a monolayer of (3-aminopropyl)trimethoxysilane instead before the deposition of gold[42]. Then, gold films with various thicknesses were deposited on the substrates under a base pressure of ~$5\times10^{-6}$ Torr at a rate of 0.2 nm/s (DISCOVERY-635, DENTON), which was followed by a lift-off process. Finally, contact electrodes of the Hall-bar structures were fabricated using the method introduced above.

Using the fabricated Hall-bar structures, the electrical properties of 2DGFs and sputtered gold films were characterized by standard low-frequency mesurements using a lock-in amplifier (SR 830, Stanford Research) and applying an alternating ($f$ = 31 Hz) current with an amplitude of 100 μA from an AC Current Souce (6221, Keithley). For the measurement of the sheet resistance (as schematically shown in Supplementary Fig. 9), the voltage between the electrodes 1 and 2 of the sample ($V_{12}$) was first measured, the sheet resistance can then be calculated as $R_S = \frac{W}{L}\frac{V_{12}}{I}$, where $W$ and $L$ are the geometrical parameters defining the Hall-bar structure and $I$ is the constant current through the Hall bar.

**Measurement of nonlinear emission spectra.** The setup for the measurement of the nonlinear emission spectrum of 2DGFs is schematically shown in Supplementary Fig. 11. For the SHG and THG measurements (Fig. 3b), laser pulses generated from a Ti:sapphire femtosecond laser (Mai Tai HP, Spectra-Physics) with a central wavelength of 1550 nm (~140-fs pulse width, 80-MHz repetition rate) was used for the excitation. A 20× objective (0.7 NA, Nikon) was used to focus the *p*-polarized incident pulses (800-mW average power) onto the samples (at a fixed incident angle of ~30° and a spot size of ~20 μm) and collect the reflective spectra of the nonlinear emission. After passing through a 950 nm short-pass filter blocking the reflected excitation laser pulses, the nonlinaer emission was directed into a charge-coupled device (CCD) camera (DS-Fi3, Nikon) and a spectrometer (Shamrock SR-750, Andor) for imaging and spectral analysis, repectively. For the SHG and MPPL measurements (Fig. 3d), laser pulses generated from a Ti:sapphire femtosecond laser (Mai Tai HP, Spectra-Physics) with a central wavelength of 800 nm (~100-fs pulse width, 80-MHz repetition rate) were used for the excitation. A 50× objective (0.8 NA, Nikon) was used to focus the *p*-polarized incident pulses (5-mW average power) onto the samples (at a fixed incident angle of ~30° and a spot size of ~1 μm) and collect the reflective spectra of the nonlinear emission. After passing through a 700 nm short-pass filter blocking the reflected excitation laser pulses, the nonlinaer emission was directed into a CCD camera and a spectrometer (QE Pro, Ocean Insight) for imaging and spectral analysis, repectively.

To avoid the impact of surface roughness on the nonlinear emission measurement, the 200-nm-thick sputtered gold film used for comparison was fabricated using a template stripping method to obtain an atomically smooth surface. Firstly, a gold film with a thickness of 200 nm was deposited onto a cleaned silicon substrate by magnetron sputtering at a base pressure of $5\times10^{-6}$ Torr at a deposition rate of 0.2 nm/s. Secondly, a droplet (10 μL) of epoxy glue (EPO-TEK 301-2) was admitted onto the gold film, followed by the placement of a cleaned glass substrate on the top. Thirdly, the structure was transferred onto a hot plate to cure the epoxy under 80 °C for 3 h, and then was slowly cooled down to room temperature. Finally, the glass substrate was detached from the silicon substrate with a gold film having an atomically smooth surface attached to it.

**Simulation of nonlinear emission.** To simulate thickness-dependent SHG and THG signals of 2DGFs, the eigen-state wave-functions and eigen-energies were first calculated using the quantum electrostatic model[6], which is based on the self-consistent solution of Schrödinger and Poisson equations. After that, a standard approach based on the perturbation theory for the intersubband transition (ISBT)-contributed nonlinearity is adopted to calculate the second- and third-harmonic susceptibilities ($\chi^{(2)}$ and $\chi^{(3)}$). Because the energy of the THG photons (2.4 eV) from 1550-nm pulse excitation is close to the interband transition (IBT) in gold, the contribution of IBT to $\chi^{(3)}$ was also calculated using the density matrix approach. To obtain the thickness-dependent SHG and THG intensities, experimental values of the incident pulse intensity as well as reflectivity at the fundamental, second- and third-harmonic wavelengths were used. Finally, all the intensities were normalized to the maxima (Supplementary Fig. 12). The simulated thickness-dependent SHG intensities with the contribution from ISBTs (Supplementary Fig. 12a) agree well with the experimental spectra (Fig. 3c, top panel). When contributions from both ISBTs (Supplementary Fig. 12b, sharp oscillations) and IBTs (Supplementary Fig. 12c, a broad peak) are considered to the simulated THG, the experimentally obtained THG thickness dependence (Fig. 3c, bottom panel) is well reproduced.

**Fabrication of nanoribbon arrays.** Nanoribbon arrays based on 2DGFs were fabricated using the local etching approach introduced above. Nanoribbon arrays based on sputtered gold films were fabricated as follows. EBL was first used to define masks of the nanoribbon array structures on a mica substrate. Secondly, to avoid the use of metallic adhesion layers (such as Cr, Ti) that can introduce significant optical loss to plasmonic structures, the exposed region of the masks was functionalized with a monolayer of (3-aminopropyl)trimethoxysilane instead before the deposition of gold[42]. Then, gold films with various thicknesses were deposited on the substrates under a base pressure of ~$5\times10^{-6}$ Torr at a rate of 0.2 nm/s (DISCOVERY-635, DENTON), which was followed by a lift-off process to obtains the nanoribbon arrays.

**Numerical simulations for nanoribbon arrays.** Transmission through gold nanoribbon arrays was numerically simulated using the finite element method (COMSOL Multiphysics software). The nanoribbon arrays were illuminated with a plane wave at normal incidence. Due to the invariance of the numerical problem in one of the directions, its dimensionality was reduced to 2D. Taking advantage of the symmetry of the structure, a unit cell of the array was modelled with periodic boundary conditions set on its sides. Perfectly matched layers were introduced at the top and bottom of the simulation domain to ensure the absence of back-reflection. The nanoribbon was set to have a rectangular cross-section with the width $w$ and thickness $t$, both of which were varied. In the case of 2DGFs the single-crystal permittivity from Olmon *et al.* was taken, while in the polycrystalline case data from Johnson and Christy was used. Additionally, to take into account additional losses related to the thickness-dependent electron scattering on the nanoribbon boundaries, the permittivities in both cases were corrected using the Fuchs theory. The refractive index of a mica substrate was taken to be 1.53.

**Acknowledgements** We thank Shuangshuang Liu from the Core Facilities, Zhejiang University School of Medicine for the technical support on MPPL imaging; and Jie Cao from the Westlake Center for Micro/Nano Fabrication for the facility support and technical assistance on EBL. This work was supported by the National Natural Science Foundation of China (Grants 62075195, 12004333 and 92250305), National Key Research and Development Project of China (Grant 2018YFB2200404), the UK EPSRC CPLAS project EP/W017075/1, Zhejiang Provincial Natural Science Foundation of China (Grant LDT23F04015F05) and Fundamental Research Funds for the Central Universities (226-2022-00147). The data access statement: all the data supporting this research are presented in full in the results section and supplementary materials.

# Supplementary Information for
# Large-area, freestanding single-crystal gold of single-nanometer thickness

*Chenxinyu Pan[1], Yuanbiao Tong[1], Haoliang Qian[2], Alexey V. Krasavin[3], Jialin Li[1], Jiajie Zhu[1], Yiyun Zhang[2], Bowen Cui[1], Zhiyong Li[1,4], Chenming Wu[1], Zhenxin Wang[1], Lufang Liu[1], Linjun Li[1,4], Xin Guo[1,4], Anatoly V. Zayats[3,\*], Limin Tong[1,5,\*] and Pan Wang[1,4,\*]*

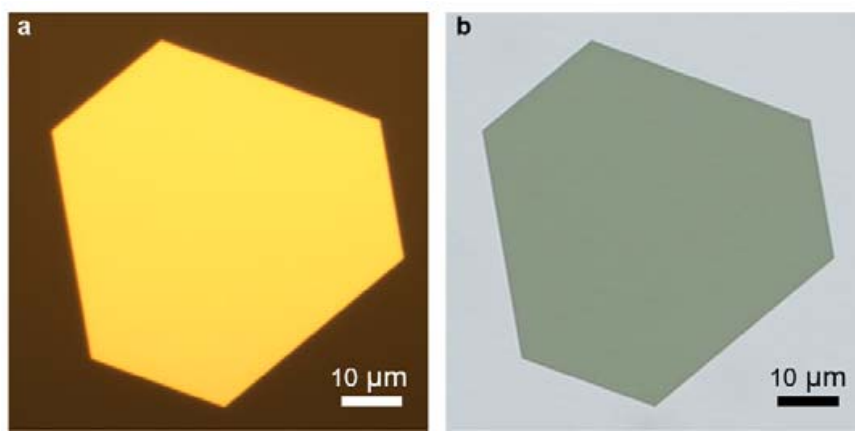

**Supplementary Fig. 1 | Optical micrographs of a typical gold flake. a,b**, Optical micrographs of a gold flake with a thickness of 30 nm taken in reflection (**a**) and transmission (**b**). Gold flakes were chemically grown on substrates using a modified wet-chemical method. By controlling the growth time, the thickness of the gold flakes can be tuned from ~10 to 100s of nanometers. Simutaneously, the lateral size of the gold flakes increses from ~10 to 100s of micrometers.

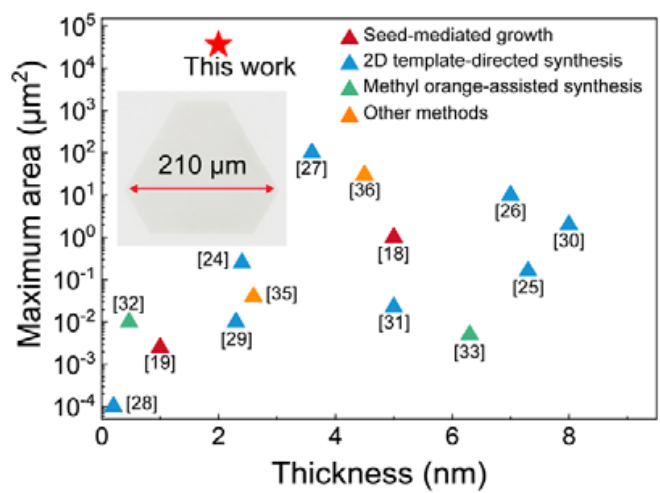

**Supplementary Fig. 2 | Comparison of area and thickness of some representative 2D gold.** Comparison of the area and thickness of a 2DGF (2 nm in thickness, 210 μm in lateral size, as shown in the inset) fabricated using the ALPE approach (red star) with counterparts obtained using other approaches (the corresponding references are given in the labels).

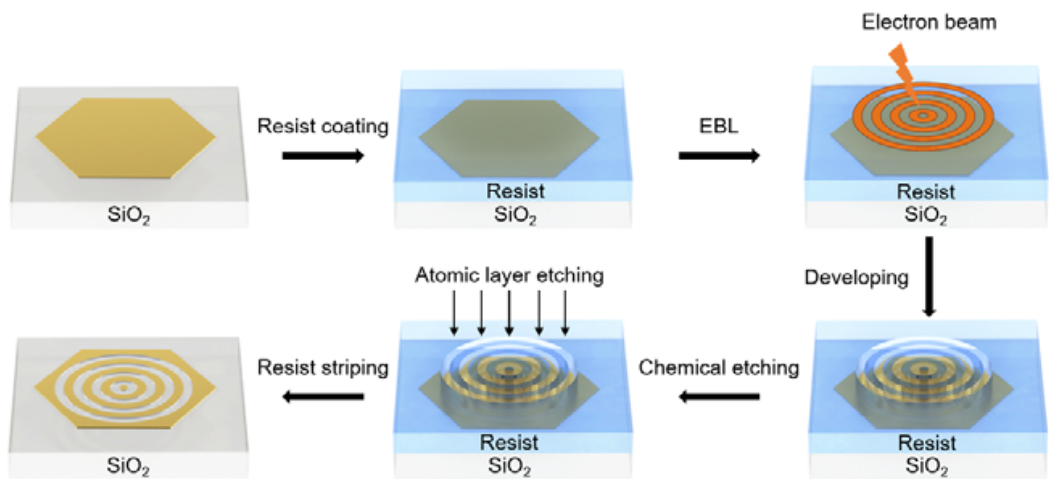

**Supplementary Fig. 3 | Schematic illustration of processes for local etching of gold flakes.** A gold flake on a substrate was first spin-coated with a 200-nm-thick resist and a 90-nm-thick conductive layer. Secondly, the resist on the gold flake was exposed with an electron beam to write the designed pattern, which was followed by a developing process to remove the exposed resist. Then, the sample was immersed into an etching solution to precisely etch the exposed gold. Finally, the resist was stripped off in acetone, and the sample was rinsed with ethanol and deionized water.

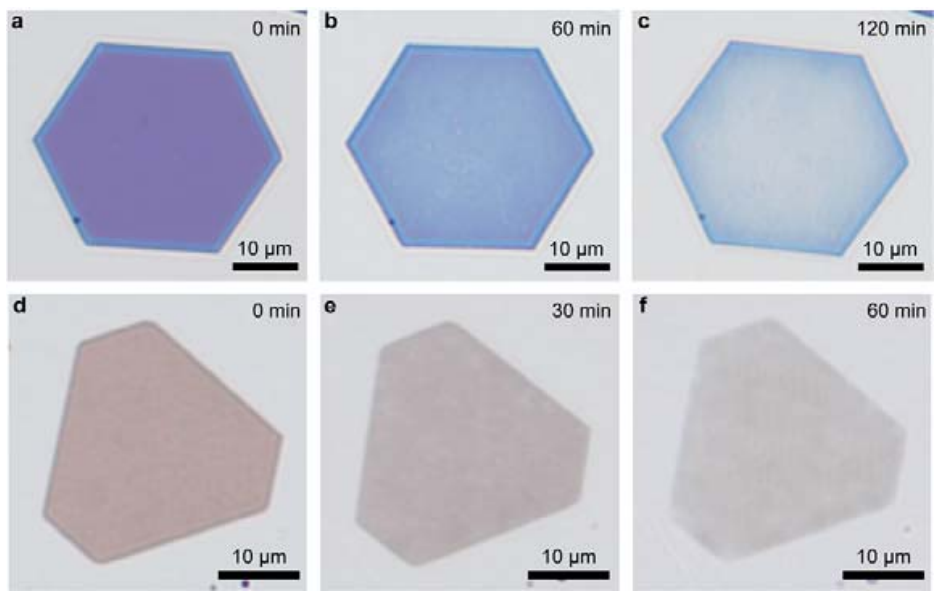

**Supplementary Fig. 4 | Fabrication of 2D silver and copper flakes using ALPE approach.** Optical transmission micrographs of silver (**a–c**) and copper (**d–f**) flakes taken at various etching times.

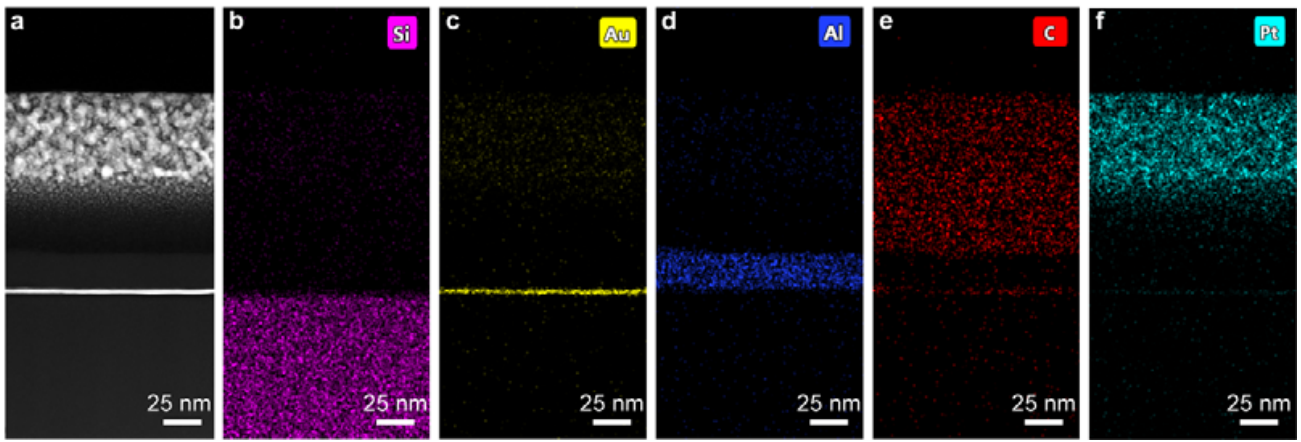

**Supplementary Fig. 5 | Elemental mapping results obtained from the cross-section of the 2DGF shown in Fig. 2d. a**, Cross-sectional TEM image of the 2DGF. **b−f**, Elemental maps of the sample presented in **a**, showing the distribution of silicon (**b**), gold (**c**), aluminum (**d**), carbon (**e**) and platinum (**f**).

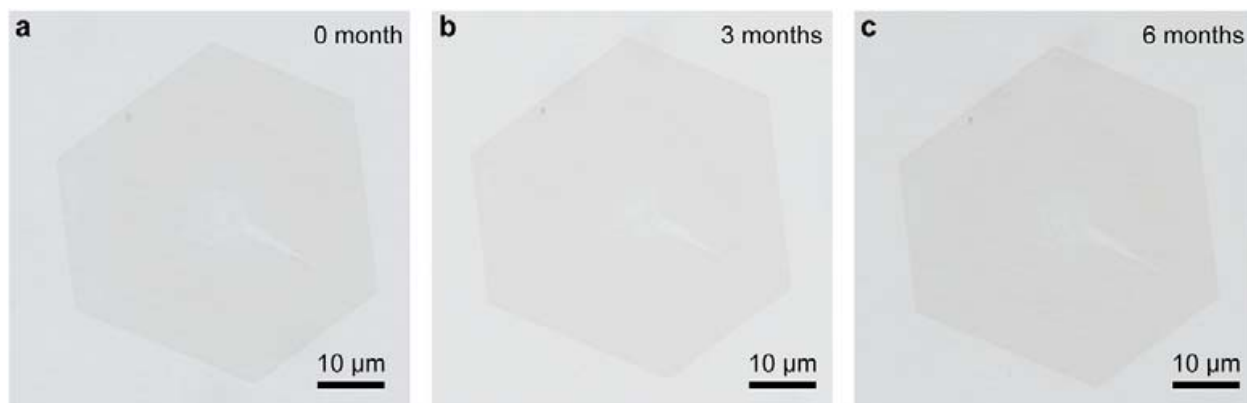

**Supplementary Fig. 6 | Long-term stability test of 2DGFs. a**, Optical micrograph of a freshly fabricated 2DGF with a thickness of 2.5 nm. **b,c**, Optical micrographs of the 2DGF taken after 3 months (**b**) and 6 months (**c**). The 2DGF was stored in ethanol to protect it from the contamination during the test. There is no observable change in the optical transmission micrographs after 6 months, indicating the excellent stability of the 2DGFs at room temperature.

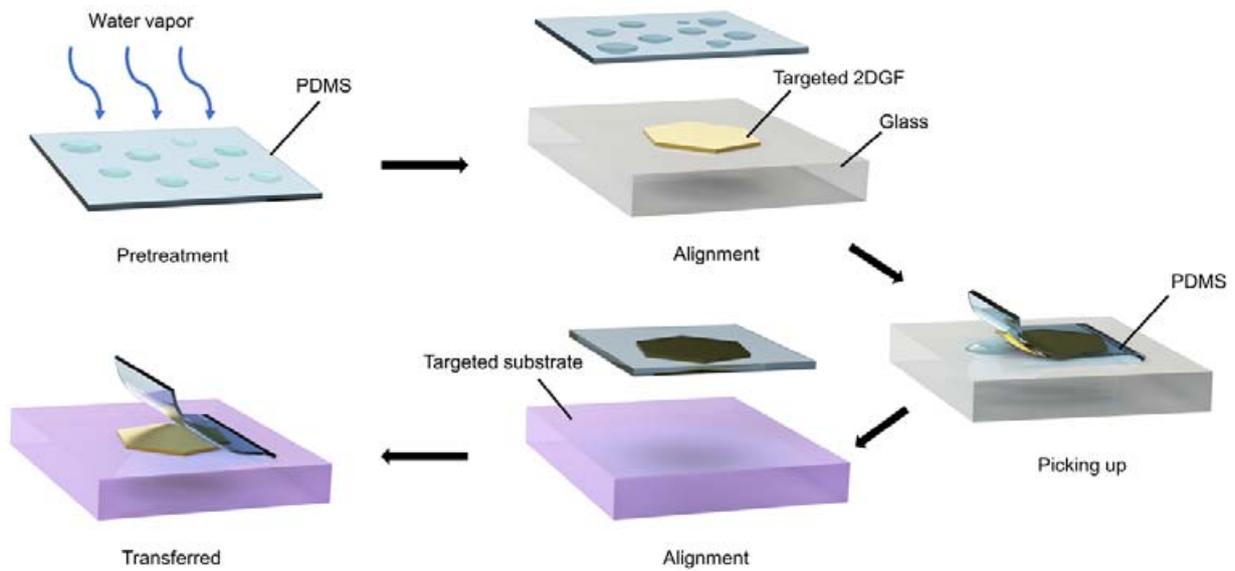

**Supplementary Fig. 7 | Schematic diagram of PDMS-assisted transfer printing of a 2DGF onto a targeted substrate.** A small patch of PDMS used for transfer printing of 2DGFs was first treated with water vapor to form numerous micrometer-scale water droplets on it. Then, under an optical microscope, the PDMS was aligned with the targeted 2DGF, and pressed onto it to pick it up. Finally, the 2DGF attached on the PDMS film was aligned and made a full contact with the targeted substrate or structure. The 2DGF was left on the substrate or structure after a careful withdrawal of the PDMS film.

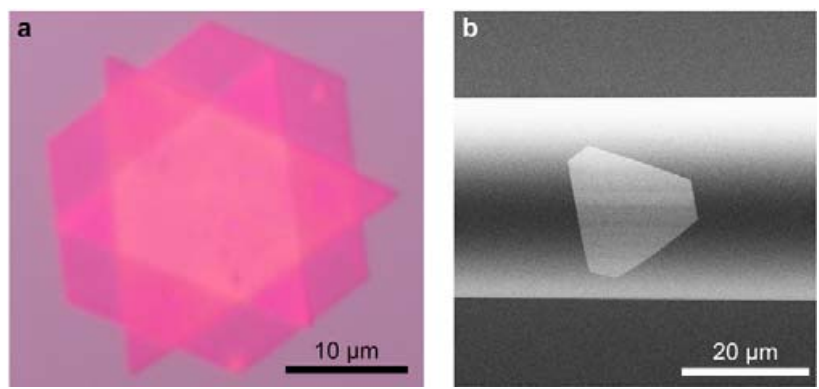

**Supplementary Fig. 8 | Transfer printing of 2DGFs. a**, Optical micrographs of a three-layer stack of 2DGFs (each ~3 nm in thickness) on a SiO$_2$/Si wafer. **b**, Scanning electron microscopy image of a 2DGF (3.1 nm in thinckness) transfer-printed onto the sidewall of a 37-μm-diameter microfiber.

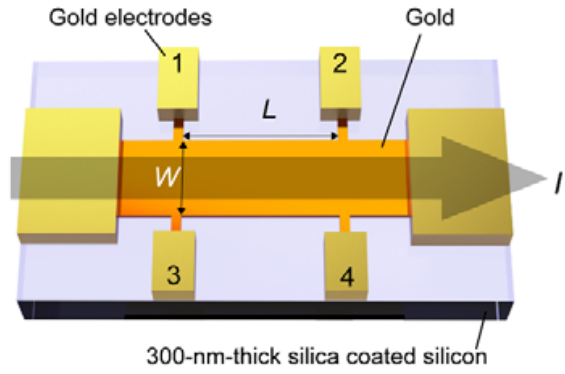

**Supplementary Fig. 9 | Schematic illustration of the Hall-bar structure used for measurements of sheet resistance.** For the measurement of sheet resistance, the voltage between electrodes 1 and 2 of the sample ($V_{12}$) was first measured, the sheet resistance can then be calculated as $R_S = \frac{W}{L}\frac{V_{12}}{I}$, where $W$ and $L$ are the geometrical parameters defining the Hall-bar structure and $I$ is the constant current through the Hall bar.

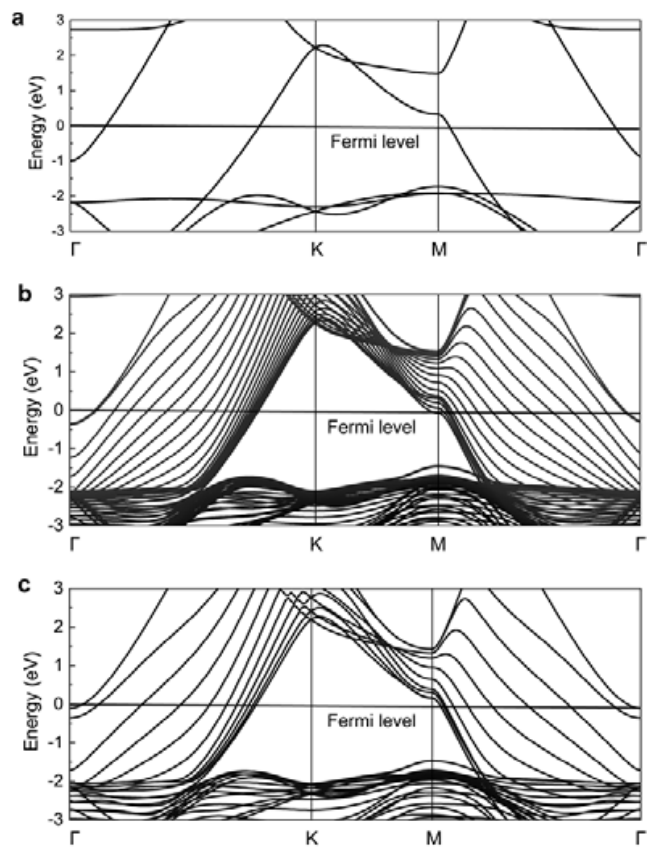

**Supplementary Fig. 10 | Thickness dependence of electronic band structures of gold. a–c,** Electron energy bands simulated for a bulk gold (**a**), a 20 atomic layer-thick gold slab (**b**) and a 10 atomic layer-thick gold slab (**c**) with a hexagonal close-packed structure. The calculations were performed using the all-electron full-potential APW + lo method as implemented in the WIEN2k code within the PBG-GGA functional. The experimental lattice constant of 4.08 Å was used for the calculations.

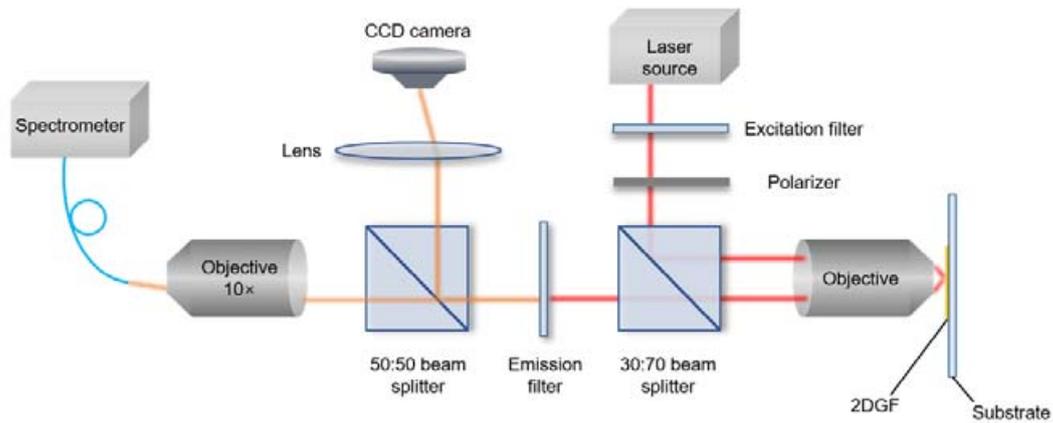

**Supplementary Fig. 11 | Setup for measurements of nonlinear emission from 2DGFs.** Collimated *p*-polarized femtosecond laser pulses used for the excitation after passing through an excitation filter were focused onto 2DGFs with an objective. The reflected nonlinear optical signal was collected by the same objective. After passing through an emission filter blocking the reflected excitation laser pulses, the nonlinaer emission was directed into a CCD camera and a spectrometer for imaging and spectral analysis, repectively.

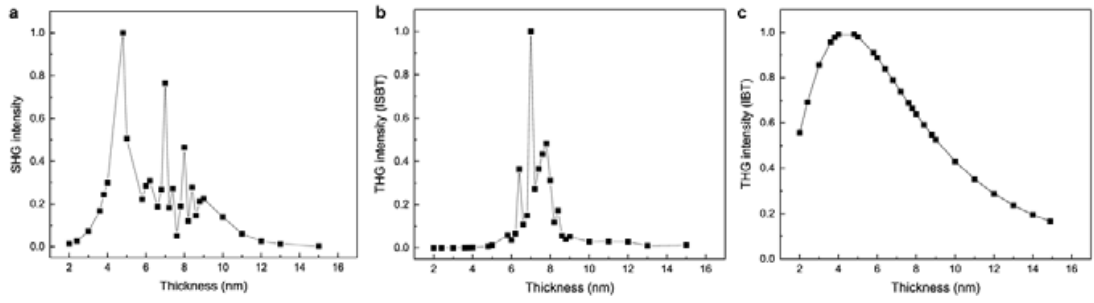

**Supplementary Fig. 12 | Simulated SHG and THG intensities. a-c**, Thickness-dependent SHG (**a**) and THG (**b**,**c**) intensities, simulated taking into account ISBTs (**a**,**b**) and IBTs (**c**). The simulated thickness-dependent SHG intensities with contribution from the ISBTs agree well with the experimental dependence (Fig. 3c, top panel). When both contributions for the THG from ISBTs (**b**) and IBTs (**c**) are considered, the experimentally obtained THG dependence is well reproduced.

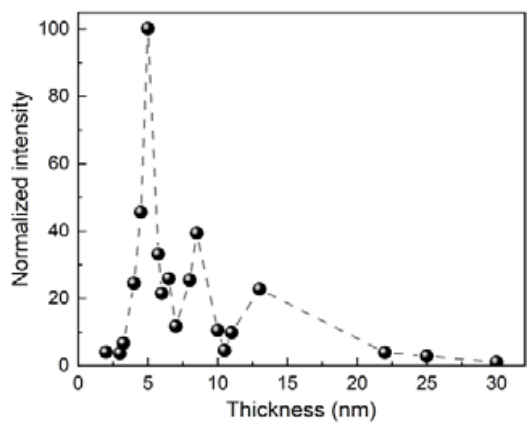

**Supplementary Fig. 13 | Thickness-dependent SHG intensities from 2DGFs excited with 800-nm laser pulses.** The dependence of SHG intensity on the thickness is extracted from the nonlinear emission spectra of 2DGFs with various thicknesses (e.g., partially shown in the inset of Fig. 3d) assuming incoherent addition of the SHG and MPPL signals, and normalized by the SHG intensity from a 30-nm-thick 2DGF.

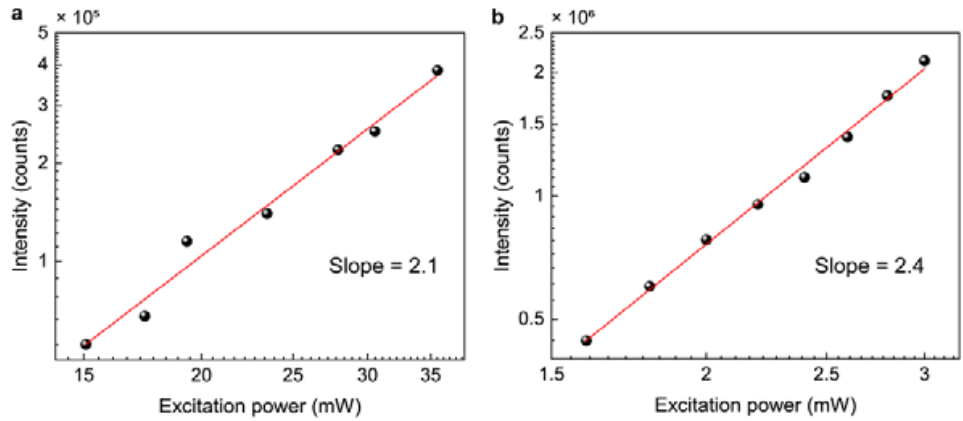

**Supplementary Fig. 14 | Determination of nonlinear orders of MPPL from gold flakes. a,b**, The dependence of spectrally-integrated intensities of MPPL from gold flakes with thicknesses of 19 (**a**) and 5 nm (**b**) on the excitation power. The spectrally-integrated MPPL intensities are extracted from the corresponding nonlinear emission spectra by assuming incoherent addition of SHG and MPPL signals. For the thickness of gold flake of 19 nm, a linear fit to the integrated MPPL intensities gives a slope of 2.1, indicating that the photoluminesce process is mainly due to two-photon photoluminesce. However, when the thickness of gold flake is decreased to, e.g., 5 nm, a slope of 2.4 is observed, showing the coexistence of two-photon and three-photon photoluminesce.

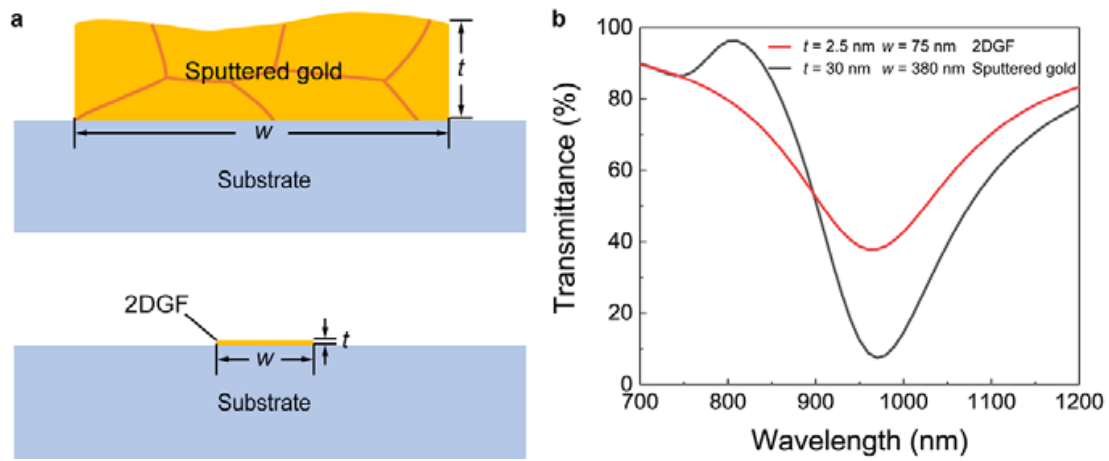

**Supplementary Fig. 15 | Comparison of plasmonic properties of nanoribbons fabricated using a 2DGF and a sputtered gold film. a**, Schematic illustrations of the nanoribbons based on a 30-nm-thick polycrystalline sputtered gold film and a 2.5-nm-thick single-crystal 2DGF. **b**, Simulated transmission spectra of nanoribbon arrays based on a 30-nm-thick sputtered gold film (black line) and a 2.5-nm-thick 2DGF (red line).

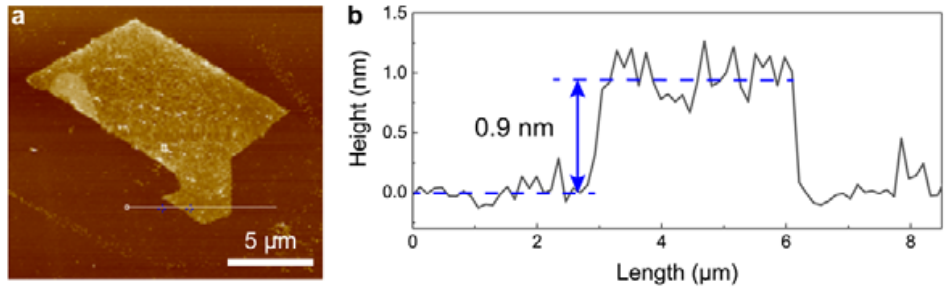

**Supplementary Fig. 16 | Sub-nanometer 2DGF. a**, AFM image of a sub-nanometer 2DGF. **b**, Line scan along the white line indicated in **a**, showing a gold thickness of ~0.9 nm, or ~5 atomic layers (cf. Fig. 2e). In principle, large area 2DGFs with thickness down to a single-atomic layer could be fabricated using the ALPE approach. However, due to the slight fluctuation in the thickness of the initial gold flake across its whole area and non-uniformity of the thickness of the native organic layer on the flake surface, it is challenging to obtain sub-nanometer 2DGFs with a large area. By further improvement of the thickness uniformity of the initial gold flake and removing the native organic layer, it is possible to push the thickness of 2DGF down to a single-atomic level.